\documentclass[12pt,preprint]{aastex}

\newcommand{\lsim}{\raise0.3ex\hbox{$<$}\kern-0.75em{\lower0.65ex\hbox{$\sim$}}}
\newcommand{\gsim}{\raise0.3ex\hbox{$>$}\kern-0.75em{\lower0.65ex\hbox{$\sim$}}}
\newcommand{\propsim}{\raise0.3ex\hbox{$\propto$}\kern-0.75em{\lower0.65ex\hbox{$\sim$}}}

\newcommand{\kms}{\rm ~km~s^{-1}}
\newcommand{\ergs}{\rm ~erg~s^{-1}}

\newcommand{\ml}{~M_\odot ~\rm yr^{-1}}

\def\EE#1{\times 10^{#1}}

\newcommand{\Mdot}{\dot M}

\begin{document}
    
\title{Inhomogeneities in type Ib/c supernovae: An inverse Compton scattering origin of the X-ray emission }

\author{C.-I. Bj\"ornsson\altaffilmark{1}}
\altaffiltext{1}{Department of Astronomy, AlbaNova University Center, Stockholm University, SE--106~91 Stockholm, Sweden.}
\email{bjornsson@astro.su.se}

\begin{abstract}
Inhomogeneities in a synchrotron source can severely affect the conclusions drawn from observations regarding the source properties. However, their presence is not always easy to establish, since several other effects can give rise to similar observed characteristics. It is argued that the recently observed broadening of the radio spectra and/or light curves in some of the type Ib/c supernovae is a direct indication of inhomogeneities. As compared to a homogeneous source, this increases the deduced velocity of the forward shock and the observed correlation between total energy and shock velocity could in part be due to a varying covering factor. The X-ray emission from at least some type Ib/c supernovae is unlikely to be synchrotron radiation from an electron distribution accelerated in a non-linear shock. Instead it is shown that the observed correlation during the first few hundred days between the radio, X-ray and bolometric luminosities indicates that the X-ray emission is inverse Compton scattering of the photospheric photons. Inhomogeneities are consistent with equipartition between electrons and magnetic fields in the optically thin synchrotron emitting regions.
\end{abstract}


\section{Introduction}

The radio emission observed from supernovae (SNe) can usually be well accounted for by synchrotron emission from a homogeneous, spherically symmetric source. In most cases though, the limited observations necessitate an assumption about the partition of energy between magnetic fields and relativistic electrons in order to deduce the size and energy content of such a source. Although the assumed sphericity finds support in VLBI observations of a few nearby radio SNe \citep[e.g.,][]{bru10,bie11}, the degree of homogeneity has been harder to estimate. 

The presence of inhomogeneities would affect the interpretation of observations in at least two ways. The low frequency part of the radio spectrum is sometimes seen to deviate from that expected from synchrotron self-absorption. This has been ascribed to free-free absorption from a thermal population of electrons \citep[e.g.,][]{che82b,wei02} and/or cooling of the relativistic electrons \citep{f/b98}. However, some of these characteristics could also be caused by an inhomogeneous source structure. Hence, care must be taken to include the possibility of inhomogeneities when analyzing such sources. In the standard model, the partition of energy between magnetic field and relativistic electrons can be determined if either the cooling frequency of the electrons or the inverse Compton scattered radiation that they give rise to can be measured. Since the inverse Compton scattered radiation is optically thin, it measures the total number of relativistic electrons independent of their degree on inhomogeneity. In contrast, the synchrotron self-absorption frequency is quite sensitive to the presence of inhomogeneities. Hence, the value deduced for the partition of energy between magnetic fields and relativistic electrons  depends on the degree of inhomogeneity.

The winds of hot, massive stars are thought to be driven by line scattering \citep{cas75}. It was soon realized that velocity perturbations made them susceptible to an instability \citep{mac79,cal80} that would lead to a clumpy wind. Several of the expected features in such a scenario have been confirmed observationally \citep{l/m99,osk08,sun11}. However, although the winds from Wolf-Rayet stars, the presumed progenitors of a large fraction of type Ib/c SNe, show signs of clumpiness, their main properties do not always conform to those expected from a line driven instability \citep{osk12}. The supernova ejecta itself may be clumpy and/or have an aspherical distribution. The polarization associated with strong lines in type Ib/c SNe has been used by \citet{tan12} to argue for a clumpy medium. Likewise, double-peaked emission line profiles in several type Ib/c SNe suggest a flattened distribution of the expanding gas \citep{maz05,mae08}. In addition,  the shock forming process could lead to inhomogeneities; for example, the contact discontinuity is likely to be Rayleigh-Taylor unstable. 

The relativistic electrons are thought to be produced by first order Fermi acceleration as they repeatedly cross the forward shock front. For a non-relativistic shock, this gives rise to an optically thin synchrotron spectral flux $F(\nu) \propto \nu^{\alpha}$ with $\alpha \approx -1/2$ \citep{bel78,b/o78} or, if synchrotron cooling is important, $\alpha \approx -1$. Many supernovae have $\alpha \sim -1$. For a supernova exploding into the wind of a red super giant, the energy density behind the shock can be high enough for synchrotron cooling to become important \citep{f/b98}. However, for Wolf-Rayet stars the density of the wind is thought to be so low that such cooling is unlikely. Inverse Compton cooling may be significant but only during the very early phases \citep[e.g., SN 2002ap;][]{b/f04}. A possible way to reconcile the observed spectra in type Ib/c SNe with a diffusive shock acceleration origin for the relativistic electrons is that the acceleration is so efficient that the whole process becomes non-linear \citep{bar99,b/e99,ell00}. This results in a concave spectrum. The steepening at low frequencies, which is relevant for the radio regime, is determined by a combination of the injection efficiency and the escape of the highest energy electrons from the acceleration process. Neither of these are well understood physically. Hence, if it could be established that the acceleration process in type Ib/c SNe is non-linear, the observed value of $\alpha \sim -1$ would strongly constrain the physical mechanisms operating in the shock region.

Inhomogeneities may provide an alternative to a non-linear acceleration process as the origin for the observed steep spectral flux. In the standard scenario they could modify the acceleration process so that a steeper distribution of electron energies results. On the other hand, if the acceleration process is associated with the contact discontinuity/the reverse shock region, the inhomogeneities may also here leave some imprint on the observed spectral flux.

A related issue concerns the origin of the X-ray emission in SNe. This is still a matter of debate \citep{d/g12}; for example, in SNe with detectable radio emission a straightforward extrapolation of the synchrotron spectrum falls substantially below the observed X-ray luminosity. \citet{c/f06} argued against a thermal origin in type Ib/c SNe. Instead they showed that with a suitable choice of parameters both the radio and the X-ray emission can be attributed to synchrotron radiation from a concave distribution of electron energies expected from a non-linear acceleration process. Another alternative is inverse Compton scattering of the photospheric photons. Such an origin for the X-ray emission in SN 2002ap is consistent with a homogeneous, spherically symmetric source \citep{b/f04}. However, SN 2002ap stands out in that its radio flux peaked early (within a few weeks). For SNe peaking at later times, the radius of the shock tends to be larger and the optical emission has started to decline. In such SNe, homogeneous and spherically symmetric models usually under predict the X-ray emission \citep{c/f06}. This conclusion hinges on the assumption that the energy is partitioned roughly equally between relativistic electrons and magnetic fields.

As mentioned above, the partition of energy and the degree of inhomogeneity are intertwined. The aim of the present paper is to discuss how the observed radio emission can be used to infer the presence of inhomogeneities and the implications it has for the X-ray emission. Focus is on acceleration at the forward shock. However, the conclusions are general enough to be applicable with only minor changes also to the case when the emission comes from the contact discontinuity/the reverse shock region. The scope of the paper is limited to type Ib/c SNe. The properties of an inhomogeneous synchrotron source are discussed in \S\,2. Special attention is given to characteristics that are unlikely to be produced by other mechanisms. The non-linear diffusive shock acceleration scenario in a supernova context is discussed in \S\,3. It is shown that it has a few implications, which may constrain its applicability. The consequences of an inhomogeneous source structure for the observed X-ray emission are considered in \S\,4, assuming the latter to be produced via inverse Compton scattering of the photospheric photons. It is shown how the combined effects of the partition of energy and inhomogeneities can be deduced from the observed radio, bolometric and X-ray luminosities.  A discussion of the results as well as a brief summary of the main conclusions follow in \S\,5. It is pointed out that the presence of inhomogeneities can have implications for the relation between ordinary Ib/c SNe  and engine-driven hypernovae as well as the basic properties of the acceleration of relativistic electrons.

\section{An inhomogeneous synchrotron source} \label{sect2}

With the increased quantity and quality of radio observations of SNe, it has become clear that spectra and/or light curves are often broader than expected for a standard homogeneous, spherically symmetric source. In the type IIb SN 1993J, \citet{f/b98} modeled the broad light curves as due to cooling of the synchrotron emitting electrons. Such an explanation requires a high density behind the shock. Since the wind velocity of the progenitor stars to type Ib/c SNe are expected to be much higher than those for the progenitor stars to type IIb SNe, the densities will be correspondingly smaller. Hence, the broad spectra/light curves in type Ib/c SNe are not likely to be caused by cooling of the same magnitude as that deduced by \citet{f/b98} in SN 1993J. 

\citet{sod05} introduced a parameter to artificially broaden the spectra in the type Ib/c SN 2003L. \citet{wei11} discussed various causes for the spectral shape in the type Ic SN 1994I. They concluded that the best fit to the observations came from free-free absorption intrinsic to the synchrotron source. The thermal electrons need to be distributed roughly as the relativistic ones, since, in order to get a good fit, the radio opacity should be the sum of synchrotron and free-free absorption. Since there are no clear indications of extrinsic free-free absorption in type Ib/c SNe, the thermal gas needs to cool after passing through the shock. With the circumstellar densities expected in type Ib/c SNe, this suggests the presence of dense clumps of thermal gas. For a clumpy gas the need for cooling sets a lower limit to their typical size \citep{chu93}. The effects on the synchrotron spectrum will be quite different depending on whether such clumps are optically thick or not to free-free absorption. Optically thick clumps produce a depression of the synchrotron spectrum without substantially changing its shape, except at the higher frequencies where the clumps become optically thin \citep{c/c06}. Hence, if the observed broadening of the radio spectra in SN 1994I is due to internal free-free absorption, it implies optically thin clumps. However, following the arguments in \citet{chu93}, the combined requirements of cooling and optically thin clumps are unlikely to be met in standard type Ib/c SNe environments. 

Below, deviations from a homogeneous, spherically symmetric source will be described in terms of the covering factor and the filling factor. The former characterizes the optically thick properties of the source while the latter accounts for its optically thin properties. One important aspect that does not enter in an analysis of an unresolved source is the geometrical distribution on the inhomogeneities.  Hence, in this paper, the concept of inhomogeneities includes also an aspherical distribution of emitting plasma. 

\subsection{Inhomogeneities with a covering factor less than unity} \label{sect2a}

The spectral distribution of the optically thick emission in an inhomogeneous source is the combination of two independent factors. Since the synchrotron source function is proportional to $B^{-1/2}$, where $B$ is the magnetic field strength, the variation of the covering factor $f_{\rm B,cov}$ with $B$ needs to be specified. Furthermore, $F(\nu)$ is proportional to the brightness temperature. At a given frequency, this, in turn, is given by the value of the Lorentz factor ($\gamma_{\rm abs}$) corresponding to a synchrotron optical depth of unity. Since $\gamma_{\rm abs} \propto (\nu / B)^{1/2}$, the second relation that needs to be specified is that between $B$ and $\nu$.  

The source is assumed to consist of a spherical shell of outer radius $R$ and thickness $R/4$. Let the observed emission along a given line of sight through the source be dominated by a magnetic field $B$ over a distance $r(B) \le R/4$. The probability to find a particular value of $B$ is taken to be $P(B) = A\,B^{-a}$ for $B_{\rm 1} < B < B_{\rm 2}$ and zero outside this range. Here, $A\, \propsim \,
B_{\rm 1}^{a-1}$ for $a>1$. The spectrum has three different regimes: For $\nu < \nu(B_{\rm 1})$, the spectrum is that for an optically thick source with magnetic field $B = B_{\rm 1}$, while for $\nu > \nu(B_{\rm 2})$, the spectrum corresponds to an optically thin source with magnetic field $B = B_{\rm 2}$. In the transition region $ \nu(B_{\rm 1}) < \nu < \nu(B_{\rm 2})$, the spectrum is given by
\begin{equation}
	F(\nu) \propto R^2 \nu^{5/2} \int \limits_{B}^{B_{\rm 2}}\frac{P(\hat{B})}{\hat{B}^{1/2}}
	{\rm d}\hat{B}.
	\label{eq:2.1}
\end{equation}
Hence, $F(\nu) \propto R^2 \nu^{5/2} B_{\rm 1}^{a-1}B^{1/2 - a}$. 

The relation $B(\nu)$ depends on the more detailed properties of the source. The distribution of Lorentz factors ($\gamma$) for the relativistic electrons is taken to be $N(\gamma) \propto \gamma^{-p}$ for $\gamma > \gamma_{\rm min}$ and $p>2$. Since observations \citep{c/f06} indicate $p \approx 3$, in order to simplify the discussion, $p=3$ is used in the following. In the transition region, $\nu^3 \propto U_{\rm e} U_{\rm B} r(B) \gamma_{\rm min}/\gamma_{\rm abs}$. Here, $U_{\rm e}$ and  $U_{\rm B} \equiv B^2 / 8\pi$ are, respectively, the energy densities in relativistic electrons and magnetic fields. Together with the expression for $\gamma_{\rm abs}$, this gives
\begin{equation}
	B \propto \frac{\nu^{7/5}}{\left(U_{\rm e}\, r(B) \gamma_{\rm min}\right)^{2/5}}.
	\label{eq:2.2}
\end{equation}
The simplest case is when only the value of $B$ varies so that $B\propto \nu^{7/5}$, which leads to\begin{equation}
	F(\nu) \propto \nu^{(16-7a)/5}.
	\label{eq:2.3}
\end{equation}	
The observations of SN 2003L by \citet{sod05} indicate that the broadening of the spectral peak is quite symmetric implying a rather flat spectrum in the transition region. This is obtained for $a \approx 16/7$. The corresponding covering factor is $f_{\rm B,cov} \approx (B/B_{\rm 1})^{1-a}$, which, together with equation (\ref{eq:2.2}), gives a covering factor varying with frequency as $f_{\nu {\rm ,cov}} \approx (\nu / \nu(B_{\rm 1}))^{-9/5}$.

The parametrization of the spectra for SN 1994I done by \citet{wei11}, can be used to estimate the frequency width of the transition region. The parameter values derived by them indicate that this width slowly decreases with time but is typically around two in the radio regime; i.e., $\nu(B_{\rm 2})/\nu(B_{\rm 1}) \approx 2$. Hence, the covering factor for the optically thin synchrotron emission (i.e., $\nu > \nu(B_{\rm 2})$) is roughly $1/4$. Although \citet{sod05} used a different parameterization to account for the broadening of the spectral peak, the implied value for the covering factor is consistent with that deduced for SN 1994I. 

\subsubsection{Implications from the light curves} \label{sect2aa}

Light curves provide complementary information to spectral data. However, an extra assumption has to be made regarding the mass loss rate and wind velocity of the progenitor star. The discussion in this section will assume that both of these are constant. Although this may be a fair approximation for the majority of type Ib/c SNe, there are possible exceptions; for example, SN 2003bg \citep{sod06} and SN 2001em \citep{c/c06}.

The three different spectral regimes translate into three different regimes for the light curve. It is seen from equation (\ref{eq:2.1}) that in the transition regime, the flux varies with time $t$ as $F(t) \propto R^2 B_{\rm 1}^{a-1}$. It is usually assumed that the energy density of the magnetic field scales with the thermal energy density behind the shock, which implies $B \propto t^{-1}$. \citet{m/m99} showed that the high shock velocities relevant for type Ib/c SNe is expected to scale with time roughly as $t^{-0.1}$; hence, $F(t) \propto t^{2.8-a}$. \citet{sod05} find for SN 2003L that $F(t) \propto t^{1.2}$ before the peak. A similar behavior can be seen in SN 1994I \citep{wei11}. This results in a value of $a$ somewhat smaller than deduced in section \ref{sect2a} from the spectral shape. It is seen from equation (\ref{eq:2.1}) that this implies a more rapid variation of the magnetic field strength with frequency than obtained with the above assumptions.

The relation $B(\nu)$ is determined by equation (\ref{eq:2.2}), which shows that a steeper variation will result from a decrease of $r(B)$ and/or $U_{\rm e}$ with increasing value of $B$. Since the covering factor decreases with the strength of the magnetic field ($a>1$), it is reasonable to expect that the same should apply for $r(B)$. On the other hand, a decreasing value of $U_{\rm e}$ with $B$ is non-trivial and constrains the mechanism(s) that accelerates the electrons and amplifies the magnetic field. An interesting case is when the total energy density/pressure of the relativistic electrons and magnetic field is constant behind the shock. Since, in the transition zone, $\nu \,\propsim \,(U_{\rm e} U_{\rm B})^{1/3}$, the conditions in the regions producing the majority of the optically thin radiation (i.e., $B = B_{\rm 2}$) correspond, roughly, to equipartition between relativistic electrons and magnetic fields. The implications of this, with respect to the origin of the observed X-ray radiation, will be discussed further in section \ref{sect4}.

A smaller value of $a$ will affect the covering factor ($f_{\rm B,cov} \approx (B/B_{\rm 1})^{1-a}$). However, as discussed above, the implied range of $B$-values in the transition regime increases so that the value of $f_{\rm B_{\rm 2},cov}$ is only marginally smaller.

\subsection{Inhomogeneities with a covering factor of unity} \label{sect2b}

The type of inhomogeneities discussed in section \ref{sect2a} is consistent with a scenario in which the wind of the progenitor and/or the ejecta of the supernova is clumpy. There is another type of inhomogeneity where the source is spherically symmetric (i.e., the covering factor is unity) but the values of the energy densities of the magnetic field and relativistic electrons both vary systematically with distance ($r$) away from the shock front. In this case, the inhomogeneity is likely due to the processes that amplify the magnetic field and accelerate the electrons. The optically thick spectral index is determined by the different variations behind the shock front of the synchrotron emissivity and absorptivity.  Since their relative variation (i.e., the source function) depends on $B$ only, the main features can be illustrated by assuming a constant density of electrons. Such a situation can be described by
\begin{equation}
 B = \left \{ 
	\begin{array}{lc}
	B_{\rm o}, \hspace{2cm}& r < r_{\rm o}\\
	B_{\rm o}\left( \frac{r}{r_{\rm o}}\right)^{\rm b}, & r_{\rm o} < r < R/4
	\label{eq:2.4}
	\end{array} 
\right. 
\end{equation}

Let $r_{\nu}$ be the value of $r$ corresponding to an optical depth of unity at the frequency $\nu$. Then for $\nu < \nu_{\rm o}$, where $r_{\nu_{\rm o}} = r_{\rm o}$, the spectrum is that for an optically thick, homogeneous source. Likewise, for frequencies such that $r_{\nu} > R/4$, corresponding to $\nu > \nu_{\rm R}$, the synchrotron emission is that for an optically thin source. Hence, in the same way as in section \ref{sect2a}, there is a transition region between the standard optically thick spectrum corresponding to a homogeneous source and the optically thin radiation.

The synchrotron optical depth $\tau$ is
\begin{equation}
	\tau \propto \int^r_0 \frac{B^{5/2}}{\nu^{7/2}} {\rm d\hat{r}}.
	\label{eq:2.5}
\end{equation}
With the use of equation (\ref{eq:2.4}), in the frequency range $\nu_{\rm o} < \nu <  \nu_{\rm R}$, this leads to
\begin{equation}
	\nu \propto \left \{ 1 + \frac{1}{5{\rm b}/2+1} \left[ \left( \frac{r_{\nu}}{r_{\rm o}}\right)^
	{5{\rm b}/2+1} -1\right] \right \}^{2/7}.
	\label{eq:2.6}
\end{equation}
Likewise, the spectral flux in the transition region is given by
\begin{eqnarray}
	F({\nu}) & \propto& \int^{r_{\nu}}_{0} \frac{B^2}{\nu} {\rm dr}\nonumber\\
	            &\propto& \nu^{-1}\left \{ 1 + \frac{1}{2{\rm b} + 1}\left[ \left (\frac{r_{\nu}}{r_{\rm o}}
	            \right)^{2{\rm b} + 1} -1\right] \right \}
	\label{eq:2.7}
\end{eqnarray}
In the limit when the contributions to both the emission and absorption increases with distance away from the shock front (i.e., b $> -2/5$), equations (\ref{eq:2.6}) and (\ref{eq:2.7}) can be combined to give
\begin{equation}
	\alpha = {\frac{5+9{\rm b}}{2+5{\rm b}}}.
	\label{eq:2.8}
\end{equation}
It is seen that for b $> 0$, the spectrum is flatter than for the homogeneous case. However, the smallest value is $\alpha = 9/5$, which, as discussed above, is too large to account for the observed spectra. In the other limit (i.e., b $< -1/2$) the main contributions to both emission and absorption come from $r \approx r_{\rm o}$. Although this can give rise to flat spectra ($\alpha \approx 0$), it can be deduced from equations (\ref{eq:2.6}) and (\ref{eq:2.7}) that the spectral range of the transition region is too small to accomodate the observations. Since, in the range $-1/2\, \lsim \,{\rm b}\, \lsim -2/5$, the spectra rise more steeply ($\alpha > 5/2$) than for the homogeneous case, it is unlikely that the type of inhomogeneities discussed here are the cause for the broadening of the observed spectra and light curves in type Ib/c SNe.

\section{X-ray emission due to synchrotron radiation from a modified electron distribution} \label{sect3}

The repeated scatterings across a shock, inherent to diffusive shock acceleration of particles,  impart momentum to the thermal gas in the upstream region. In order to conserve momentum and energy, this pre-acceleration of the upstream gas implies that energy/particles must escape from the flow/acceleration process \citep{b/e99}. When the injection efficiency is low, the effects on the upstream gas are negligible and the flow is characterized only by the shock compression ratio. At some critical value of the injection efficiency, the upstream flow is significantly affected by the acceleration process and particle escape needs to be taken into account. In addition to the shock compression ratio ($r_{\rm sub}$), the flow is now characterized also by the total compression ratio ($r_{\rm tot}$). In this case, the flow is determined not only by the injection efficiency but also the momentum above which particles escape the acceleration process. Likewise, the energy distribution of the accelerated particles is modified by the changing flow characteristics. As discussed by \citet{ell00}, the standard energy distribution (i.e., $p = 2$) is roughly valid in an intermediate energy range but steepens at low energies ($p_{\rm low} > 2$) and flattens at high energies ($p_{\rm high} < 2$). 

The value of $p_{\rm high}$ is rather model independent as long as the acceleration is in the non-linear regime. The flow upstream of the shock is then determined by the momentum input from the high energy particles. A self consistent description of this process gives  $p_{\rm high} \approx 1.5$ \citep{b/e99}. The value of $p_{\rm low}$ is related to $r_{\rm sub}$ in the same way as in standard shock acceleration. However, in non-linear shock acceleration the value of $r_{\rm sub}$ results from a self consistent calculation of the escaping energy/particles. Hence, the value of $p_{\rm low}$ is more model dependent than that of $p_{\rm high}$, since it is sensitive to the details of both the injection process and the escape of particles. Neither of these processes are well understood physically and are normally treated by introducing two or more parameters. Furthermore, the calculated energy distribution is that for the momentum carrying particles (i.e., ions). It is usually assumed that the diffusion length depends only on energy (e.g., Bohm diffusion), so that the energy distribution of electrons can be taken to be similar to that of the ions. The type Ib/c SNe studied by \citet{c/f06} have $p_{\rm low}\approx 3$ (i.e., $\alpha \approx -1$) with a rather small spread, which corresponds to $r_{\rm sub}\approx 2.5$. In contrast to the generic value $p_{\rm high} \approx 1.5$, this value of  $p_{\rm low} $ constrains the physics of non-linear shock acceleration.

The energy density of the magnetic field as well as that of the low energy part of the electron distribution (i.e., the energy density at the injection momentum) are normally assumed to be some (constant) fraction of the thermal energy density behind the shock. Although the total compression ratio is larger for non-linear as compared to standard shocks (i.e., $r_{\rm tot} > 4$), the thermal energy density behind the shock is smaller. Hence, the implied mass loss rate of the progenitor star increases correspondingly. For $r_{\rm sub}\approx 2.5$, standard shock relations show that the thermal energy density behind the shock decreases by a factor $r_{\rm tot}$. Furthermore, the influx of energy decreases by a factor $r_{\rm tot}/4$, decreasing the column density of magnetic field and relativistic electrons by the same amount. This increases the implied energy density of the magnetic field by a factor $(r_{\rm tot}/4)^{8/19}$ ($p_{\rm low} = 3 $). Taken together, these effects increase the needed mass loss rate of the progenitor star by a factor $4 (r_{\rm tot}/4)^{27/19}$. The numerical calculations done by \citet{c/f06} to fit the observations of SN 1994I have $r_{\rm tot} \approx 46$. The implied mass loss rate of the progenitor star is then more than a factor $10^2$ larger for non-linear shock acceleration as compared to standard shock acceleration.

It could be argued that the energy density of the magnetic field scales with the total energy density behind the shock rather than the thermal one. In this case $U_{\rm B} \approx r_{\rm tot} U^{\rm low}_{\rm e}$, where $U^{\rm low}_{\rm e}$ is the energy density of the relativistic electrons corresponding to the $p_{\rm low}$-part of the energy distribution. The implied value for the magnetic energy density would increase by a factor $\approx r_{\rm tot}^{8/19}$ ($p_{\rm low} = 3 $). The needed increase of the mass loss rate of the progenitor is then $\approx (r_{\rm tot}^{2}/4)^{8/19}$, which is roughly a factor $10$ lower than for the case in which the energy density of the magnetic field is assumed to scale with the thermal energy density.

Adiabatic heating may not be the only process which determines the temperature structure upstream of the shock. Additional heating processes will decrease the compression ahead of the shock and, hence, decrease the value of $r_{\rm tot}$ \citep{b/e99}. It is seen from the calculations by \citet{c/f06} that $p_{\rm high} \approx 1.5$ is needed to simultaneous fit the radio and X-ray emission from SN 1994I; basically, a flattening at higher energies of the electron distribution is required to offset the steepening of the spectrum due to synchrotron cooling. In order to keep the generic $p_{\rm high} \approx 1.5$, the value of $r_{\rm tot}$ cannot decrease too much. The estimates of the mass loss rates of the progenitor stars in \citet{c/f06} was done using radio data interpreted within a standard shock scenario. Since their deduced values are already quite high, assuming the presence of non-linear shock acceleration may well require prohibitively large mass loss rates.

Since the electrons radiating in the X-ray regime are cooling, the time evolution of the X-ray luminosity ($L_{\rm x}$) from a homogeneous, spherically symmetric source is \citep[e.g.,][]{b/f04} 
\begin{equation}
	L_{\rm x} \propto v_{\rm sh}^3 B^{(p-2)/2},
	\label{eq:3.1}
\end{equation}
where $v_{\rm sh}$ is the shock velocity. Assuming a constant mass loss rate and that the energy density of the magnetic field scales with the thermal energy density behind the shock, $B \propto t^{-1}$. Furthermore, in self-similar solutions for the shock propagation \citep{che82a}, the high velocities relevant for type Ib/c SNe lead to $v_{\rm sh} \propto t^{-0.12}$ \citep{m/m99}. With $p \approx p_{\rm high} \approx 1.5$, one finds $L_{\rm x}\, \propsim\, t^{-0.11}$. Hence, the X-ray luminosity is expected to decline slowly. This is consistent with the numerical results in \citet{c/f06}; for example, in their calculations the X-ray luminosity of SN 1994I declines by less than a factor $2$ from $t = 100$\,days to $t = 1000$\,days. Except for an apparent increase in SN 1994I during its initial phase, the observed X-ray variability in SNe 1994I, 2001ig, and 2003bg \citep{c/f06} suggests a decrease considerably faster than predicted from equation (\ref{eq:3.1}). This is in line with the observed X-ray variability in other type Ib/c SNe \citep[e.g.,][]{sod08}. It was pointed out by \citet{c/f06} that the X-ray emission in SN 1994I on days $82$ and $\sim 2500$ could not both be due to synchrotron radiation from electrons accelerated in a non-linear shock.

Taken together, the issues raised above show that invoking non-linear shock acceleration to explain the X-ray emission meets with some complications. Therefore, in the next section, the merits of inverse Compton scattering of the photospheric photons as the origin of the X-ray emission are considered.

\section{X-ray radiation from inverse Compton scattering of the photospheric photons}\label{sect4}

During the first few weeks after explosion, inverse Compton scattering of photospheric photons can contribute significantly to the X-ray emission in type Ib/c SNe.  At later times, such an origin for the observed X-ray emission is more model dependent; for example, a homogeneous, spherically symmetric source, in which the energy densities in magnetic fields and relativistic electrons are in equipartition, falls short by at least an order of magnitude \citep{c/f06}. However, in light of the discussion in section \ref{sect2}, an inverse Compton origin of the X-ray radiation may still be viable. In this section, the assumption of homogeneity is relaxed with the aim to elucidate how this affects the deduced source properties. The type of inhomogeneities considered is limited to those discussed in section \ref{sect2a}.
 
Let $L_{\rm r}$ and $L_{\rm bol}$ denote the optically thin radio luminosity and the bolometric luminosity of the supernova, respectively. Since the optically thin radio spectral index is close to $\alpha = -1$, $L_{\rm r} \propto \nu F({\nu})$ is roughly independent of frequency. In order to facilitate the discussion it will be assumed that the distribution of relativistic electrons has the same form throughout the source although its amplitude may vary.
 
Since the inverse Compton scattered radiation is optically thin, it is unaffected by inhomogeneities and depends only on the total number of relativistic electrons. With $<U_{\rm e}>$ denoting the average energy density of relativistic electrons, the inverse Compton scattered radiation can be written $L_{\rm x} / L_{\rm bol} = (4/3)\gamma^2\, \tau(\gamma)$, where $\tau(\gamma)$ is the scattering optical depth of electrons with Lorentz factor $\gamma$. Hence, 
\begin{eqnarray}
	\gamma^2\, \tau(\gamma) & = & \frac{<U_{\rm e}> \gamma_{\rm min}}{mc^2}\frac{p-2}{p-1}
	\left(\frac{\gamma}{\gamma_{\rm min}}\right)^{3-p} \sigma_{\rm T} \frac{R}{4}\nonumber  \\
	                          & = &  \frac{<U_{\rm e}> \gamma_{\rm min}}{8\,mc^2} \sigma_{\rm T} R
	                          \hspace{4cm}{\rm for} \;p= 3,
	\label{eq:4.1}
\end{eqnarray}
where $\sigma_{\rm T}$ is the Thomson cross-section. It has been assumed that the thickness of the shell is $R/4$ and $p = 3$ corresponds to $\alpha = -1$. The average energy density of electrons can then be expressed as
\begin{equation}
	<U_{\rm e}> = \frac{6\, mc^2}{\gamma_{\rm min}\sigma_{\rm T}R}\frac{L_{\rm x}}{L_{\rm bol}}
	\label{eq:4.2}
\end{equation}

The synchrotron emitting regions that contribute most of the optically thin radio emission is assumed to correspond to a volume filling factor $f_{\rm B,vol}$. In order to simplify the notation, in this section $B$ corresponds to $B_{\rm 2}$ in section \ref{sect2a} so that, for example, $f_{\rm B,vol} \approx f_{\rm B_{\rm 2},cov} r(B_{\rm 2})/(R/4)$. The energy density of the magnetic field in these regions is denoted by $U_{\rm B}$. Since synchrotron radiation can be regarded as inverse Compton scattering of the virtual photons due to the magnetic field,
\begin{equation}
	\frac{L_{\rm r}}{L_{\rm x}} = \frac{f_{\rm B,vol}\, U_{\rm B}\, \delta}{U_{\rm ph}}.
	\label{eq:4.3}
\end{equation}
Here, $U_{\rm ph} \equiv L_{\rm bol} / 4\pi R^{2} c$ is the energy density of photons due to the photospheric emission and $\delta \equiv U_{\rm e} / <U_{\rm e}>$, where $U_{\rm e}$ is the energy density of electrons in the synchrotron emitting regions. The partition of the energy densities of relativistic electrons and magnetic fields in the regions responsible for the optically thin radio emission is then given by
\begin{equation}
	\frac{U_{\rm e}}{U_{\rm B}} = \frac{24 \pi mc^3}{\sigma_{\rm T}} f_{\rm B,vol}\, \delta^2
	\frac{R}{\gamma_{\rm min}}\frac{1}{L_{\rm r}}\left(\frac{L_{\rm x}}{L_{\rm bol}}\right)^2
	\label{eq:4.4}
\end{equation}

\subsection{Mass-loss rates and shock velocities}\label{sect4a}

In order to minimize the mass-loss rate of the progenitor star it is often assumed that all the available electrons are injected into the acceleration process. This allows a determination of both the mass-loss rate and the value of $\gamma_{\rm min}$. With the assumption that a fraction $\epsilon_{\rm e}$ of the thermal energy density behind the shock is converted into relativistic electrons,
\begin{equation}
	\gamma_{\rm min} = \frac{9}{32}\frac{p-2}{p-1}\epsilon_{\rm e} \frac{m_{\rm p}}{m}
	\left(\frac{v_{\rm sh}}{c}\right)^2 \mu_{\rm e},
	\label{eq:4.5}
\end{equation}
where $m_{\rm p}$ and $\mu_{\rm e}$ denote, respectively, the proton mass and the mean molecular weight of the electron and, for $p = 3$, this can be written
\begin{equation}
	\gamma_{\rm min} = 1.9 \times 10 (3\,\epsilon_{\rm e})\frac{\mu_{\rm e}}{2}
	\left(\frac{3\,v_{\rm sh}}{c}\right)^2.
	\label{eq:4.6}
\end{equation}
For a progenitor wind devoid of hydrogen and complete ionization behind the shock $\mu_{\rm e} \approx 2$.

The mass-loss rate ($\Mdot$) is related to $<U_{\rm e}>$ through
\begin{equation}
	<U_{\rm e}> = \frac{9\,\epsilon_{\rm e}}{32\pi}\frac{\Mdot}{v_{\rm w}}
	\left(\frac{v_{\rm sh}}{R}\right)^2,
	\label{eq:4.7}
\end{equation}
where $v_{\rm w}$ is the wind velocity of the progenitor. With the use of equation (\ref{eq:4.2}), this implies
\begin{equation}
	\frac{\Mdot_{\rm -5}}{v_{\rm w,3}} = 1.8\EE{2} \left(\frac{c}{3v_{\rm sh}}\right)^3
	\frac{t_{\rm 1}}{\left(3\epsilon_{\rm e}\right)^2}\frac{L_{\rm x}}{L_{\rm bol}}.
	\label{eq:4.7a}
\end{equation}
Here $\Mdot_{\rm -5} \equiv 1\EE{-5} \ml$, $v_{\rm w,3} \equiv v_{\rm w}/(10^3 \kms)$, and the observing time $t$ is written as $t_{\rm 1} \equiv t/10$\,days. Furthermore, the expression for $\gamma_{\rm min}$ in equation (\ref{eq:4.6}) has been used.

In an inhomogeneous source, the shock radius is not directly related to the radiating volume. Since only a fraction (i.e.,$f_{\rm B,vol}$) of the volume behind the shock front contributes, the shock velocity needs to be larger than for the homogeneous case. Again, in order to simplify the notation, $\nu_{\rm m} \equiv \nu(B_{\rm 2})$ will be used. Standard synchrotron theory then gives for a spherical source,
\begin{eqnarray}
	B & \propto & \frac{\nu_{\rm m}f_{\rm B,cov}^{2/17}}{F(\nu_{\rm m})^{2/17}} 
	\left (\frac{U_{\rm B}}{U_{\rm e} (\gamma_{\rm abs}) \Delta}\right)^{4/17} 
	\label{eq:4.8}\\
	R & \propto & \frac{F(\nu_{\rm m})^{8/17}} {\nu_{\rm m}f_{\rm B,cov}^{8/17}}
	\left(\frac{U_{\rm B}}{U_{\rm e} (\gamma_{\rm abs}) \Delta}\right)^{1/17}
	\label{eq:4.9},
\end {eqnarray}
where $U_{\rm e}(\gamma_{\rm abs})$ is the energy density of relativistic electrons radiating at frequencies larger than $\nu_{\rm m}$ and $\Delta \equiv 4\,r(B)/R$. Using the expression for $<U_{\rm e}>$ in equation (\ref{eq:4.2}), equations (\ref{eq:4.8}) and (\ref{eq:4.9}) can be written
\begin{eqnarray}
	B & \propto & \nu_{\rm m}^{7/5}\left(\frac{L_{\rm bol}}{L_{\rm x}} \frac{1}{\Delta\, \delta}
	\right)^{2/5} \label{eq:4.10}\\
	R & \propto & \frac{F(\nu_{\rm m})^{1/2}}{\nu_{\rm m}^{9/10} f_{\rm B,cov}^{1/2}}
	\left(\frac{L_{\rm bol}}{L_{\rm x}} \frac{1}{\Delta\, \delta}\right)^{1/10}. 
	\label{eq:4.11}
\end{eqnarray}
Since $L_{\rm r} \approx \nu_{\rm m} F(\nu_{\rm m})$, it is seen that equation (\ref{eq:4.4}) is implied by equations (\ref{eq:4.10}) and(\ref{eq:4.11}).

\subsection{Comparing SN 2003L to SN 2002ap}\label{sect4b}

Although several of the unknown parameters characterizing the inhomogeneities affect the value of the radius and, hence, the shock velocity, it is seen from equation (\ref{eq:4.11}) that the main effect is due to the covering factor so that $v_{\rm sh}\, \propsim\, f_{\rm B,cov}^{-1/2}$. As discussed in section \ref{sect2a}, the value of $f_{\rm B,cov}$ can be estimated from the broadening of the radio spectra and/or light curves. The radio luminosities of type Ib/c SNe span a rather wide range with SN 2002ap (low) and SN 2003L (high) as extreme cases. Although the observations of SN 2002ap are not as extensive as those of SN 2003L, there are no indications of spectral broadening. Since $f_{\rm B,cov} \sim 1/4$ is deduced for SN 2003L in section \ref{sect2a}, its actual shock velocity would then be a factor $\sim\,2$ larger than deduced for the homogeneous case. This makes the velocity almost equal to that derived in \citet{c/f06} for SN 2002ap.

The volume filling factor of the optically thin synchrotron emission can be estimated from equation (\ref{eq:4.4}),
\begin{equation}
	\frac{U_{\rm e}}{U_{\rm B}} = 2.8 \times 10^2\, f_{\rm B,vol}\, \delta^2 \frac{t_{\rm 40}}
	{3\epsilon_{\rm e}} \frac{c}{3v_{\rm sh}}\left(\frac{L_{\rm x,L}}{L_{\rm bol,L}}\right)^2 
	\frac{1}{L_{\rm r,L}}.
	\label{eq:4.12}
\end{equation}
The observed values are normalized to those given by \citet{sod05} for SN 2003L at $t \approx 40~$days (i.e., $L_{\rm x} = 9.2\times 10^{39} {\ergs}, L_{\rm r} = 0.074 \times L_{\rm x} = 6.8 \times 10^{38} {\ergs}$, and $L_{\rm bol} = 1.5\times 10^{42} {\ergs}$). The value of $f_{\rm B,vol}$ depends on the relative distribution of magnetic fields and relativistic electrons and, hence, it is harder to estimate than that of $f_{\rm B,cov}$ from the available observations. However, as was discussed in section \ref{sect2a}, the combination of light curves and spectra can be used to constrain the source properties. 

One way to account for the radio light curves in SN 2003L is to assume an anti-correlation between the energy densities of electrons and magnetic fields. In order to illustrate how this can be used to deduce the value of $f_{\rm B,vol}$, consider the special case when their sum is constant throughout the source. This implies that the optically thin synchrotron radiation comes mainly from regions characterized by rough equipartition between electrons and magnetic fields. Since $\delta \approx 1/2$ for equipartition, one finds from equation (\ref{eq:4.12}) that $f_{\rm B,vol} \approx 1/70$ for $\epsilon_{\rm e} =1/3$. Furthermore, with $f_{\rm B,cov} \approx 1/4$, the thickness of the emission region is $\Delta \approx 1/17$.

Comparing SN 2002ap to SN 2003L, one finds from the values given in \citet{c/f06} that at $t \approx 4~$days (i.e., a factor $10$ earlier than for SN 2003L), ($L_{\rm x}/L_{\rm bol})^2 /L_{\rm r}$ is a factor two smaller. Hence, the RHS of equation (\ref{eq:4.12}) is a factor 20 smaller for SN 2002ap as compared to SN 2003L. Assuming, for example, equipartition between relativistic electrons and magnetic fields in the optically thin emission regions then implies $f_{\rm B,vol} \delta^2 \approx 0.1$. The reason that \citet{b/f04} found that the observations of SN 2002ap are consistent with a homogeneous source in equipartition is that they assumed cooling to be important. Since, in this case $p \approx 2$, the energy density in relativistic electrons varies only logarithmically with $\gamma_{\rm min}$. On the other hand, for $p=3$, the value of $\gamma_{\rm min}$ is important since then $U_{\rm e} \propto 1/\gamma_{\rm min}$. Hence, assuming cooling to be negligible requires either $U_{\rm e}/U_{\rm B} \approx 10$ or an inhomogeneous emission region also in SN 2002ap. However, in the latter case, the inhomogeneities are likely to be less pronounced than in SN 2003L; in particular, observations suggest $f_{\rm B,cov} \approx 1$ in SN 2002ap.

The mass-loss rate derived from equation (\ref{eq:4.7a}) for the progenitor star to SN 2003L on $t = 40$\,days is 
\begin{equation}
         \frac{\Mdot_{\rm -5}}{v_{\rm w,3}} = 4.4\left(\frac{c}{3v_{\rm sh}}\right)^3
	\frac{1}{\left(3\epsilon_{\rm e}\right)^2}.
	\label{eq:4.12a}
\end{equation}
This is roughly a factor $10^3$ larger than the corresponding mass-loss rate for SN 2002ap. Hence, interpreting the X-ray emission as inverse Compton scattering of photospheric photons leads to the conclusion that the main difference between SN 2003L and SN 2002ap is the much larger mass-loss rate associated with the former supernova.

\subsection{Implications for other SNe of type Ib/c }\label{sect4c}

Although SN 2003L and SN 2002ap have rather different observed properties, the value of $(L_{\rm x}/L_{\rm bol})^2 /L_{\rm r}$ differ only by a factor two. Figure\,\ref{fig1} shows that variations in this quantity are quite small also amongst other type Ib/c SNe; in particular, in comparison with the large variations of the radio luminosity. The exception is the late time observations of SN 1994I. The observations at $t \sim 2500~$days give a very large value and are not included in Figure\,\ref{fig1}. This tendency of an increasing value of $(L_{\rm x}/L_{\rm bol})^2 /L_{\rm r}$ can be seen already at $t \approx 82~$days. A possible explanation is an additional source of X-ray emission, which is supported by the fact that the X-ray luminosity of SN 1994I actually increased between days $t \approx 52$ and $t \approx 82$. It should also be noticed that a certain amount of scatter in $(L_{\rm x}/L_{\rm bol})^2 /L_{\rm r}$ is expected since: (1) Different energy bands were used to calculate $L_{\rm x}$. (2) In the derivation of equation (\ref{eq:4.12}) $p=3$ was assumed. The radio spectra indicate some variation around this value and \citet{c/f06} used the observed spectral index to deduce $L_{\rm r}$ at $5~$GHz.

Rewriting equation (\ref{eq:4.12}) or using equations (\ref{eq:4.10}) and (\ref{eq:4.11}) leads to
\begin{equation}
	\left(\frac{L_{\rm x}}{L_{\rm bol}}\right)^2 \frac{1}{L_{\rm r}} \propto \frac{U_{\rm e} \epsilon_{\rm 	e}v_{\rm sh}}{U_{\rm B}} \frac{1}{t f_{\rm B,vol}\, \delta^2}
	\label{eq:4.13}
\end{equation}
Although the observations are still rather sparse, the small scatter in Figure\,\ref{fig1} shows that they are consistent with $L_{\rm r}\, \propsim\, (L_{\rm x}/L_{\rm bol})^2$ during the first few hundred days of interaction between the supernova shockwave and the circumstellar medium. This is as expected for a scenario in which a substantial contribution to the X-ray emission comes from inverse Compton scattering of photospheric photons. 

For a homogeneous source the partition of energy between relativistic electrons and magnetic fields is directly obtained from $(L_{\rm x}/L_{\rm bol})^2 /L_{\rm r}$. Although its value is not so straightforward to deduce for an inhomogeneous source, equation (\ref{eq:4.13}) indicates that  one may expect the value of $(L_{\rm x}/L_{\rm bol})^2 /L_{\rm r}$ to decrease with time for an individual source. Unfortunately, X-ray observations of individual sources are not yet extended enough for such a comparison to be made. However, it is seen in Figure\,\ref{fig2} that for the SNe discussed in this paper, the value of $(L_{\rm x}/L_{\rm bol})^2 /L_{\rm r}$ vary only weakly, if at all, with time. Unless these SNe happened to be observed at particular times, this suggests that the same conclusion holds also for individual sources. Hence, an inverse Compton scenario for the X-ray emission implies not only an inhomogeneous source structure but also that variation of the inhomogeneities in type Ib/c SNe to some extent cancels variation in $t$. 

To some extent the properties of the inhomogeneities can be constrained from the detailed radio observations. The optically thin radio luminosity $L_{\rm r} \propto \gamma_{\rm min}U_{\rm e}U_{\rm B}R^3 f_{\rm B,vol}$. With the standard assumption that $U_{\rm e}\propto U_{\rm B} \propto 1/t^2$, this leads to $L_{\rm r} \propto v_{\rm sh}^5 f_{\rm B,vol}/t$. With $v_{\rm sh}\propto t^{-0.12}$ \citep{m/m99}, the observed time variation of $L_{\rm r}$ \citep {c/f06} indicates $f_{\rm B,vol} \sim$\,constant (SN 2002ap is an exception; its radio luminosity decreases considerably slower than those for the rest of the type Ib/c SNe). With a roughly constant value of $f_{\rm B,vol}$, equation (\ref{eq:4.13}) suggests that the main variation with time is a decreasing value of $\delta$, i.e., the average energy density of relativistic electrons ($<U_{\rm e}> $) decreases slower with time than that in the synchrotron emitting regions ($U_{\rm e}$).

\section{Discussion and conclusions}\label{sect5}

Compared to a homogeneous synchrotron source, the presence of inhomogeneities implies a higher velocity for the forward shock. The observed radio spectral index ($\alpha \approx -1$) makes the injection energy of the relativistic electrons/low energy cut-off ($\gamma_{\rm min}$) an important parameter, since the total number of relativistic electrons $\propto \gamma_{\rm min}^{-2}$.  Together with the assumption that all the electrons are injected into the acceleration process, inhomogeneities lead to a lower mass-loss rate for the progenitor star. On the other hand, the higher velocity of the forward shock increases the total energy (and/or decreases the mass of the ejecta). Hence, if the physical parameters of a source are deduced using a homogeneous model, a varying covering factor ($f_{\rm B,cov}$) would result in an apparent correlation between total energy and shock velocity. This could, to some extent, account for the difference between ordinary type Ib/c SNe and those suggested to be powered by a central engine.

\begin{figure}
\epsscale{0.95}
\plotone{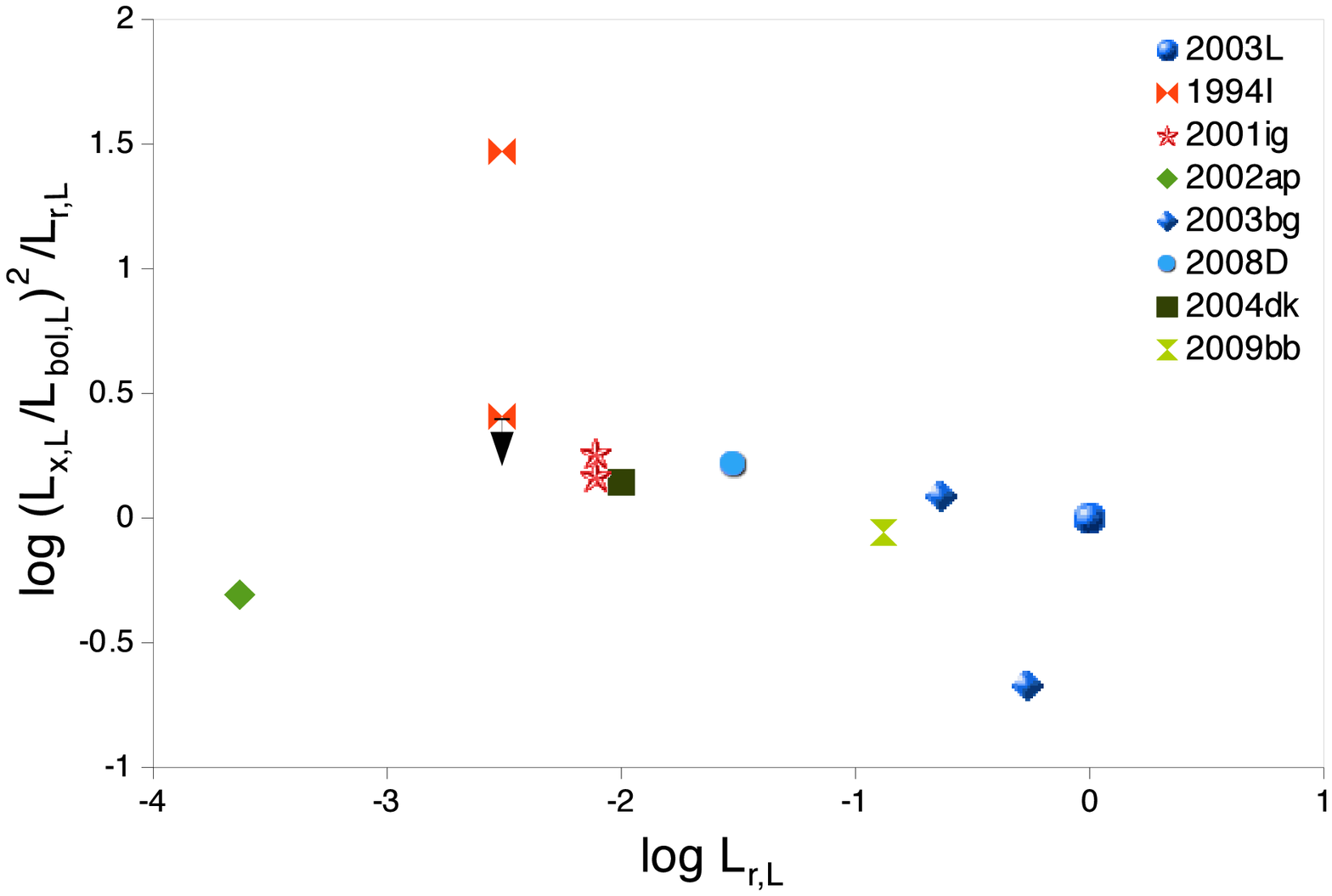}
\caption{The observed values of the inverse Compton parameter $(L_{\rm x}/L_{\rm bol})^2 / L_{\rm r}$  versus radio luminosity $L_{\rm r}$. The values have been normalized to those of SN 2003L. The lack of a correlation is evidence for an inverse Compton origin of the X-ray emission (see text). The values for 2003L, 1994I, 2001ig, 2002ap, and 2003bg are taken from the compilation of \citet{c/f06}. The other type Ib/c SNe shown are 2008D \citep{sod08}, 2004dk \citep{wel12,dro11}, and 2009bb \citep{sod10,pig11}. The downward pointing arrow for SN 1994I corresponds to the upper limit of the X-ray emission on day 52.
\label{fig1}} 
\end{figure}

\begin{figure}
\plotone{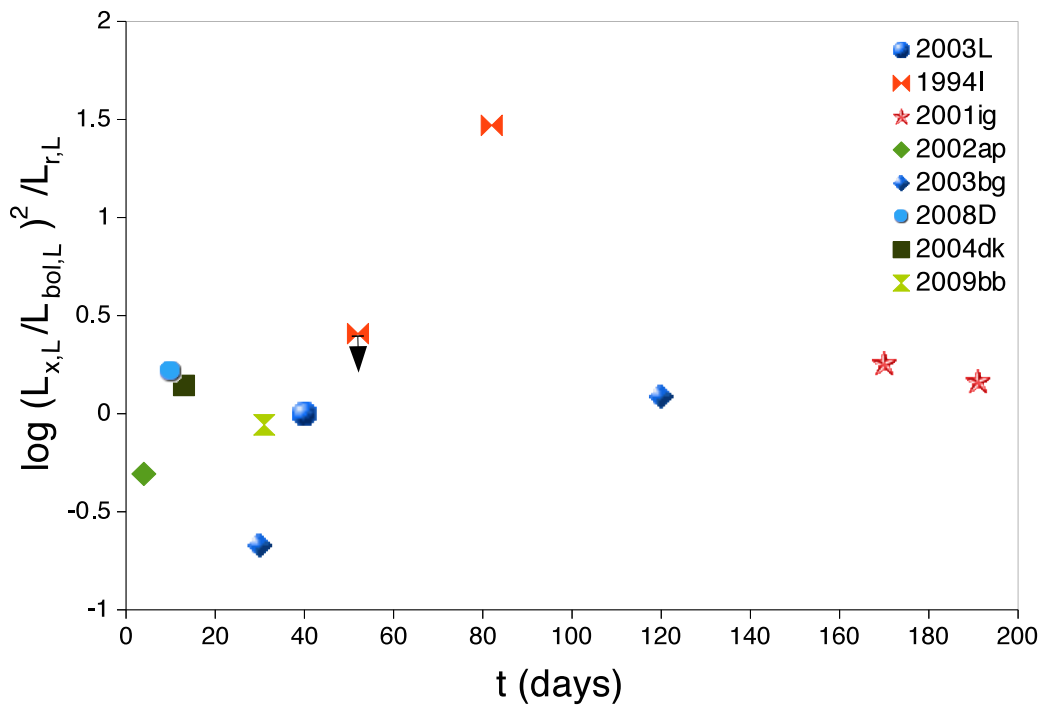}
\caption{The observed values of the inverse Compton parameter $(L_{\rm x}/L_{\rm bol})^2 / L_{\rm r}$  versus time $t$ after explosion for the same SNe as in Figure\,\ref{fig1}. The weak variation with time constrains the properties of the inhomogeneities (see text). The downward pointing arrow for SN 1994I corresponds to the upper limit of the X-ray emission on day 52.
\label{fig2}}
\end{figure}

The value of $f_{\rm B,cov}$ deduced from the broadening of the spectra and/or light curves gives a maximum value, since it is possible that not the whole source contributes to the optically thick synchrotron radiation. With this maximum value of $f_{\rm B,cov}$, it was shown that the velocity of the shock in SN 2003L would be similar to that in SN 2002ap. However, smaller values of $f_{\rm B,cov}$ and, hence, higher shock velocities cannot be excluded. The SNe in Figure\,\ref{fig1} are all ordinary type Ib/c except for SN 2009bb, which is suggested to have an engine by \citet{sod10}. It is seen that SN 2009bb follows the same relation as that of the ordinary SNe; in particular, its radio luminosity does not stand out in comparison with those of the most luminous of the ordinary type Ib/c SNe. The main difference is instead its low value of the synchrotron self-absorption frequency ($\nu_{\rm m}$) at comparable epochs. For an electron energy distribution with $p=3$, the optically thin radio luminosity $L_{\rm r} \propto  \nu_{\rm m}^3 f_{\rm B,cov}$. It is therefore possible that the low value of the synchrotron self-absorption frequency is compensated for by a larger covering factor. This then necessitates an even higher shock velocity for the ordinary type Ib/c SNe than that deduced from the comparison between SN 2003L and SN 2002ap and would, at the same time, further reduce the difference between "ordinary" and "engine-driven" type Ib/c SNe. 

The high shock velocities also suggest a high value for $\gamma_{\rm min}$. In an inverse Compton scenario, it may therefore be possible to observe the low energy cut-off  in the soft X-ray range. The rather hard X-ray spectrum ($\alpha \approx -1/2$) observed by \citet{sod05} for SN 2003L  could be caused by such an effect. Unfortunately, the low number of detected photons prevents a more detailed comparison. 

The observed radio spectra are steeper than expected from diffusive shock acceleration in a standard, non-relativistic shock. Inhomogeneities can influence several of the assumptions underlying the standard result. The fraction of electrons that scatters back across the shock front as well as their average momentum gain depend on the scattering properties of the medium. The assumption of isotropic scattering may not be applicable for an inhomogeneous source structure. Furthermore, it is normally assumed that the velocity of the scattering centers downstream of the shock does not deviate too much from the average advection velocity. However, if the inhomogeneities are associated with regions of higher densities, they would be decelerated less at the shock front, i.e., have a higher velocity relative to the shock front than the rest of the medium. In cases when they take part in the scattering of the particles, the average momentum gain per shock crossing would go down, which would steepen the spectrum.

In conclusion, the main points of the present paper can be summarized as follows:

1) The presence of inhomogeneities in a synchrotron source will affect several of the main conclusions drawn from observations; for example, the observed correlation between total energy and shock velocity could, at least in part, be due to a varying covering factor.

2) It is often hard to distinguish the effects of inhomogeneities from those of other mechanisms such as free-free absorption and cooling. It is argued that the broadening of the radio spectra and/or light curves is best understood as due to inhomogeneities; in particular, for most of the type Ib/c SNe in which cooling is likely to be negligible.

3) Models, which attribute the X-ray emission in type Ib/c SNe to synchrotron radiation from a concave electron energy distribution produced in a non-linear shock, have a few implications that could limit their applicability; for example, the mass loss rate of the progenitor star needs to be substantially larger than deduced using a standard synchrotron model and the expected decline of the X-ray emission is slower than observed in many SNe.

4) The observed correlation during the first few hundred days in type Ib/c SNe between the radio, X-ray and bolometric luminosities is the one expected from an inverse Compton scattering origin of the X-ray emission. An inhomogeneous source structure is consistent with equipartition conditions in the regions giving rise to the optically thin synchrotron emission.

It is clear that inverse Compton scattering can only contribute significantly to the X-ray emission as long as the supernova is reasonably luminous. The X-ray emission during later phases needs to be attributed to another mechanism. Although free-free emission from a homogeneous source is not a viable option in many cases, this would change with sufficiently large density inhomogeneities \citep[e.g.,][]{chu93}. The radio emission is proportional to the product of the energy densities in the magnetic field and relativistic electrons. If both of these scale with the thermal energy density, one expects the radio emission to correlate with the free-free component of the X-ray emission. Hence, inhomogeneities due to a clumpy medium would preferentially increase both of these emission components, while leaving the inverse Compton emission from either the thermal or relativistic electrons unchanged. 

The lack of broadening of radio spectra and/or light curves cannot be taken as evidence for a homogeneous, spherically symmetric source. For an unresolved source, assuming such a model results in a minimum value for the shock velocity. An independent indication for inhomogeneities is, therefore, a higher shock velocity deduced from observations at other wavelength ranges. When the shock velocity can be directly observed, unresolved inhomogeneities will give rise to a radio brightness temperature lower than expected. The low brightness temperature in the type IIb SN 1993J was interpreted by \citet{f/b98} as due to cooling. Alternatively, it could be caused by inhomogeneities, which would then lower the deduced strength of the magnetic field.

\bigskip
Thanks are due to an anonymous referee whose detailed and critical reading did much to improve the paper.

\clearpage

\clearpage

\begin{thebibliography}{}

    \bibitem[Baring et al.(1999)]{bar99} Baring, M.G., et al., 1999, \apj, 513, 311
    
    \bibitem[Bell(1978)]{bel78} Bell, A.R., 1978, \mnras, 182, 147

    \bibitem[Berezhko \& Ellison(1999)]{b/e99} Berezhko, E.G., \& Ellison. D.C.,
    1999, \apj, 526, 385
    
    \bibitem[Bietenholz et al.(2011)]{bie11} Bietenholz, M., et al., 2011, arXiv:1103.1783

    \bibitem[Bj\"{o}rnsson \& Fransson(2004)]{b/f04} Bj\"{o}rnsson, C.-I., \& Fransson, C., 
     2004, \apj, 605, 823
     
     \bibitem[Blandford \& Ostriker(1978)]{b/o78} Blandford, R.D., \& Ostriker, J.P., 1978, 
     \apj, 221, L29
     
    \bibitem[Brunthaler et al.(2010)]{bru10} Brunthaler, A., et al., 2010, \aap, 516, 27
    
    \bibitem[Carlberg(1980)]{cal80} Carlberg, R.G., 1980, \apj, 241, 1131
    
    \bibitem[Castor et al.(1975)]{cas75} Castor, J.J., Abbott, D.C., \& Klein, R.I., 1975,
    \apj, 195, 157
     
    \bibitem[Chevalier(1982a)]{che82a} Chevalier, R.A., 1982, \apj, 258, 790
    
    \bibitem[Chevalier(1982b)]{che82b} Chevalier, R.A., 1982, \apj, 259, 302
    
    \bibitem[Chevalier \& Fransson(2006)]{c/f06} Chevalier, R.A., \& Fransson, C.,
    2006, \apj, 651, 381
    
    \bibitem[Chugai(1993)]{chu93} Chugai, N.N., 1993, \apj, 414, L101
    
    \bibitem[Chugai \& Chevalier(2006)]{c/c06} Chugai, N.N., \& Chevalier, R.A., 
    2006, \apj, 641, 1051
    
    \bibitem[Drout et al.(2011)]{dro11} Drout, M.R., et al., 2011, \apj, 741, 97

    \bibitem[Dwarkadas \& Gruszko(2012)]{d/g12} Dwarkadas, V.V., \& Gruszko, J., 2012, \mnras,
    419, 1515
    
    \bibitem[Ellison et al.(2000)]{ell00} Ellison, D.C., Berezhko, E.G., \& Baring, M.G.,
    2000, \apj, 540, 292
    
    \bibitem[Fransson \& Bj\"{o}rnsson(1998)]{f/b98} Fransson, C., \& Bj\"{o}rnsson, C.-I., 
    1998, \apj, 509, 861
    
    \bibitem[Lepine \& Moffat(1999)]{l/m99} Lepine, S., \& Moffat, A.F.J., 1999, \apj, 514, 909
       
    \bibitem[MacGregor et al.(1979)]{mac79} MacGregor, K.B., Hartmann, L., \& Raymond, J.C.,
    1979, \apj, 231, 514
    
    \bibitem[Maeda et al.(2008)]{mae08} Maeda, K., et al., 2008, Science, 319, 1220 
    
    \bibitem[Matzner \& McKee(1999)]{m/m99} Matzner, C.D., \& McKee, C.F., 
    1999, \apj, 510, 379
    
    \bibitem[Mazzali et al.(2005)]{maz05} Mazzali, P.A., et al., 2005, Science, 308, 1284
        
    \bibitem[Oskinowa et al.(2008)]{osk08} Oskinova, L.M., Hamann, W.-R., \& Feldmeier, A., 
    2008, in Clumping in Hot Star Winds, Eds, W.-R. Hamann, A. Feldmeier \& L.M. Oskinova 
    (Potsdam: Univ.-Verl.), 203
    
    \bibitem[Oskinova et al.(2012)]{osk12} Oskinova, L.M., et al., 2012, arXiv:1202.1525
    
    \bibitem[Pignata et al.(2011)]{pig11} Pignata, G., et al., 2011, \apj, 728, 14

    \bibitem[Soderberg et al.(2005)]{sod05} Soderberg, A.M., et al., 2005, \apj, 621, 908
    
    \bibitem[Soderberg et al.(2006)]{sod06} Soderberg, A.M., et al., 2006, \apj, 651, 1005
    
    \bibitem[Soderberg et al.(2008)]{sod08} Soderberg, A.M., et al., 2008, \nat, 453, 469
     
    \bibitem[Soderberg et al.(2010)]{sod10} Soderberg, A.M., et al., 2010, \nat, 463, 513
    
    \bibitem[Sundqvist et al.(2011)]{sun11} Sundqvist, J.O., Owocki, S.P., \& Puls, J.,
    2011, arXiv:1110.0485

    \bibitem[Tanaka et al.(2012)]{tan12} Tanaka, M., et al., 2012, arXiv:1205.4111
    
    \bibitem[Weiler et al.(2002)]{wei02} Weiler, K.W., et al., 2002, \araa, 40, 387
    
    \bibitem[Weiler et al.(2011)]{wei11} Weiler K.W., et al., 2011, \apj, 740, 79
    
    \bibitem[Wellons et al.(2012)]{wel12} Wellons, S., Soderberg, A.M., \& Chevalier, R.A., 2005, 
    \apj, 621, 908
    
       
\end{thebibliography}
\end{document}